\documentclass[pre,reprint,twocolumn,superscriptaddress]{revtex4-2}

\pdfoutput=1

\usepackage[caption=false]{subfig}
\usepackage{amsmath}
\usepackage{physics}
\usepackage{amssymb}
\usepackage{siunitx}
\usepackage{comment}
\usepackage{bm}
\usepackage{hyperref}
\usepackage[dvipsnames]{xcolor}
\hypersetup{
setpagesize=false,
 bookmarksnumbered=true,%
 bookmarksopen=true,%
 colorlinks=true,%
 linkcolor=Blue,
 citecolor=Blue,
 urlcolor = Blue,
}
\usepackage{graphicx, color}

\newcommand{\fig}[1]{Fig.~\ref{#1}}

\newcommand{\RM}[1]{\mathrm{#1}}

\usepackage{pdfpages} 
\usepackage{pgffor} 

\makeatletter
\AtBeginDocument{\let\LS@rot\@undefined}
\makeatother

\def\supplementfilename{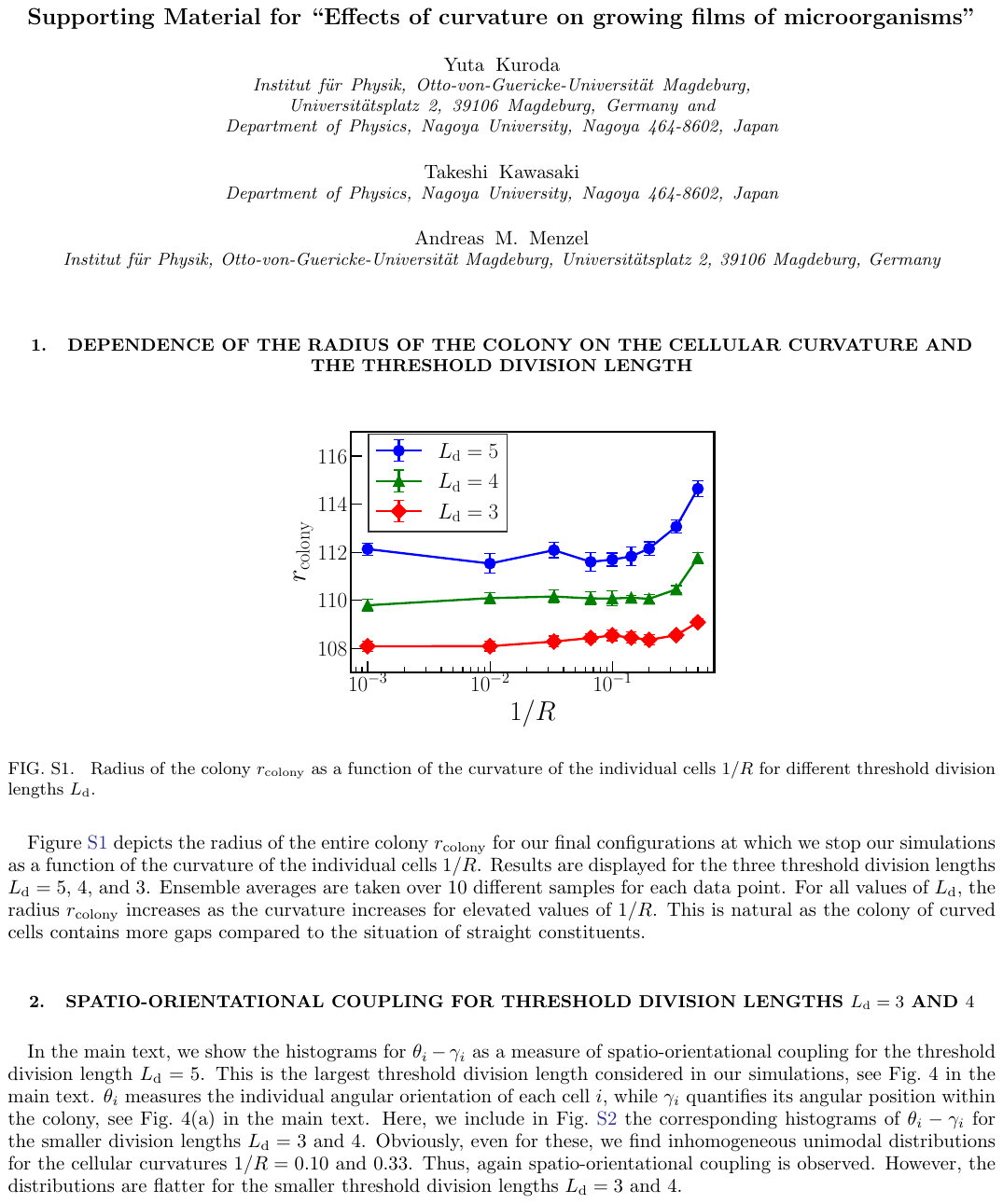}

\pdfximage{\supplementfilename}
\def\numbersupplementpages{\the\pdflastximagepages}

\newif\ifarXiv
\arXivtrue 

\begin{document}


\title{Effects of curvature on growing films of microorganisms}

\author{Yuta Kuroda}
\email{kuroda@r.phys.nagoya-u.ac.jp}
\affiliation{Institut f{\"u}r Physik, 
Otto-von-Guericke-Universit\"at Magdeburg, Universit\"atsplatz 2, 39106 Magdeburg, Germany}
\affiliation{Department of Physics, Nagoya University, Nagoya 464-8602, Japan}

\author{Takeshi Kawasaki}
\email{kawasaki@r.phys.nagoya-u.ac.jp}
\affiliation{Department of Physics, Nagoya University, Nagoya 464-8602, Japan}

\author{Andreas M.\ Menzel}
\email{a.menzel@ovgu.de}
\affiliation{Institut f{\"u}r Physik, 
Otto-von-Guericke-Universit\"at Magdeburg, Universit\"atsplatz 2, 39106 Magdeburg, Germany}

\date{\today}

\begin{abstract}
To provide insight into the basic properties of emerging structures when bacteria or other microorganisms conquer surfaces, it is crucial to analyze their growth behavior during the formation of thin films. In this regard, many theoretical studies focus on the behavior of elongating straight objects. 
They repel each other through volume exclusion and divide into two halves when reaching a certain threshold length. 
However, in reality, hardly any object of a certain elongation is perfectly straight. Therefore, we here study the consequences of the curvature of individuals on the growth of colonies and thin active films. 
This individual curvature, so far hardly considered, turns out to qualitatively affect the overall growth behavior of the colony. Particularly, strings of stacked curved cells emerge that show branched structures, while the size of orientationally ordered domains in the colony is significantly decreased. 
Furthermore, we identify emergent spatio-orientational coupling that is not observed in colonies of straight cells.  
Our results are important for a fundamental understanding of the interaction and spreading of microorganisms on surfaces, with implications for medical applications and bioengineering.
\end{abstract}

\maketitle
\begin{figure*}[hbt!]
\centering
  \includegraphics[width=17.0cm]{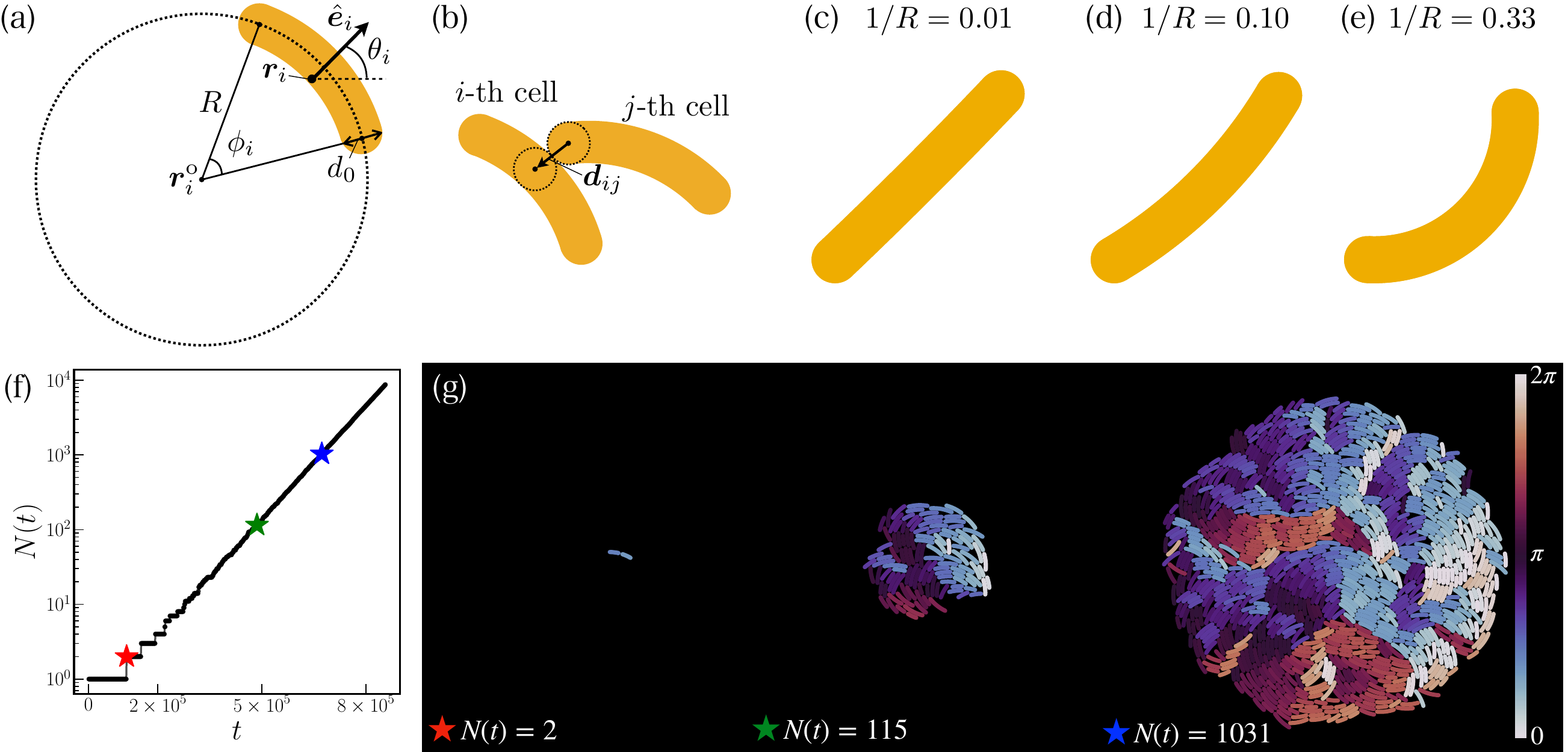}
  \caption{\label{fig1}Illustration of our curved model microorganisms. (a)~Each microorganism $i$ is represented by a curved spherocylinder of thickness $d_0$ following an arc of radius $R$ and segment angle $\phi_i$. We characterize its direction by the unit vector $\hat{\bm e}_i$ pointing in the direction from the 
  center-of-mass position $\bm r_i$ radially outward, normal to the arc. The angle $\theta_i$ measures the orientation of $\hat{\bm e}_i$ in the two-dimensional plane. $\bm{r}_i^\RM{o}$ denotes the center-of-arc position for cell $i$. 
  (b)~An illustration of calculating the mutual forces of steric interaction between two curved microorganisms. For any two cells $i$ and $j$ in contact, these forces are determined from the overlap of the nearest inner tangent circles of the curved spherocylinders, with center-to-center vector $\bm d_{ij}$, using the Hertzian elastic model. (c--d) Illustrative examples of elongated cells for curvatures of (c) $1/R=0.01$, (d) $1/R=0.10$, and (e) $1/R=0.33$. 
  (f)~A typical time evolution of the total number of cells $N(t)$ of curvature $1/R=0.10$ and threshold division length $L_\RM{d}=5$.
  The star symbols indicate configurations of $N(t)=2$, $115$, and $1031$, as depicted in panel (g).
  Color represents the orientation angle $\theta_i$ of the cells.
  }
\end{figure*}
Bacteria and other microorganisms show various modes of exploring their environment. Examples are gliding or crawling on surfaces~\cite{wu2011self, nakane2013helical, wadhwa2022bacterial}, swimming through fluids~\cite{elgeti2015physics, lauga2016bacterial, park2017multifunctional, hoell2019multi}, or simple expansion by individual growth in a crowd while pushing neighbors away to other locations~\cite{you2018geometry, tjhung2020analogies, hallatschek2023proliferating,  wittmann2023collective}. 
Understanding the behavior of these microorganisms is significant in exploring how they coexist with us in our environment.
For example, this concerns the formation of biofilms~\cite{mazza2016physics, maier2021physical} growing on teeth or prostheses. 

From a basic, theoretical, and generic perspective, the formation of such films has been investigated in detail. Particularly, this concerns the formation of monolayers. To provide a theoretical perspective as general as possible, the microorganisms are typically represented by straight spherocylinders~\cite{ghosh2015mechanically, farrell2017mechanical, acemel2018computer, you2018geometry}, that is, cylinders with hemispherical caps on their ends. These spherocylinders grow in length and divide into two halves when reaching a certain threshold elongation. In this way, basic features of bacterial growth are mimicked. Mutual interactions via volume exclusion between the spherocylinders are introduced by soft potentials, frequently employing the Hertzian model~\cite{landau1986theory}. 
Accordingly, spherocylinders push neighboring cylinders away as they expand, increasing the area of the film.
Modifications, improvements, and specifications were introduced by considering, for instance, the threshold length upon which division occurs~\cite{farrell2017mechanical}, growth rates that depend on the bacterial length~\cite{volfson2008biomechanical}, the role of nutrient supply~\cite{matsushita1990diffusion, ghosh2015mechanically, farrell2017mechanical}, or growth rates that depend on the local mechanical pressure~\cite{winkle2017modeling}. Corresponding experimental observations and measurements set the benchmark for such theoretical studies~\cite{acemel2018computer, you2018geometry}. 

In analogy to ordering phenomena on molecular scales in liquid crystals~\cite{degennes1993physics}, orientational alignment between nearby microscopic objects can be defined in expanding colonies of elongated microorganisms~\cite{doostmohammadi2016defect}. Spherocylinders provide a nematic axis that does not distinguish between head or tail. The amount of alignment of one axis with those of the surrounding objects sets the local degree of orientational order. Typically, for larger films, one no longer observes a global orientational order over the entire area.  Instead, domains of alignment form that can be distinguished by obvious boundaries~\cite{volfson2008biomechanical, acemel2018computer, you2018geometry}. There, the local axis of alignment shows a pronounced jump. The dependence of the typical domain size on the extension of the film and the threshold length at which division occurs have been analyzed in detail for straight microorganisms, both experimentally and theoretically~\cite{acemel2018computer, you2018geometry}. 

During all these investigations, the major focus was placed on straight objects. Yet, in reality, hardly any object of such size is perfectly straight. Apart from rather random curvature due to imperfection, various bacteria maintain a curved shape by construction~\cite{ausmees2003bacterial, mukhopadhyay2009curvature, schuech2019motile, taylor2019bent}, including \textit{V.\ cholerae}~\cite{Cabben2009, Bartlett2017} and \textit{C.\ crescentus}~\cite{Persat2014}.
It was argued that the cell shape is affected by trading off efficiencies of swimming, chemotaxis, and cell construction~\cite{schuech2019motile}. 
Moreover, curved bacterial cells can be generated artificially and on purpose by growing them in external electric fields~\cite{rajnicek1994electric} or under confinement in microchambers~\cite{takeuchi2005controlling}. 
For self-propelling flexible bacteria that can acquire curved shapes, deviations from straight elongation impair overall orientational order~\cite{janulevicius2010cell}.
Generally, ordering phenomena differ for curved objects, as is illustratively obvious on a phenomenological level, but also already on a molecular scale in the context of banana-shaped liquid crystals~\cite{jakli2018physics, brand2005tetrahedratic, Rico12020Science}. Therefore, here we investigate the consequences arising from the curved nature of the individual constituents in growing flat films of microorganisms. 

To this end, we introduce a theoretical model to study the growth of the films. 
It is based on discrete growing mesoscopic curved objects that divide into two halves when their arc length reaches a certain threshold. 
We illustrate the effects of curvature of the individual microorganisms on the structure of the resulting films. 
Particular attention is paid to the average size of ordered domains and how it depends on the curvature of the individuals. Moreover, the influence of the threshold length of cell division is investigated. 
We also discuss the statistical properties of the orientation of curved cells and report a spatio-orientational coupling that we do not observe for straight objects. 
In addition, we briefly address different structures emerging for growing curved objects.

\section*{Materials and Methods}
\begin{figure*}[t]
\centering
  \includegraphics[width=13cm]{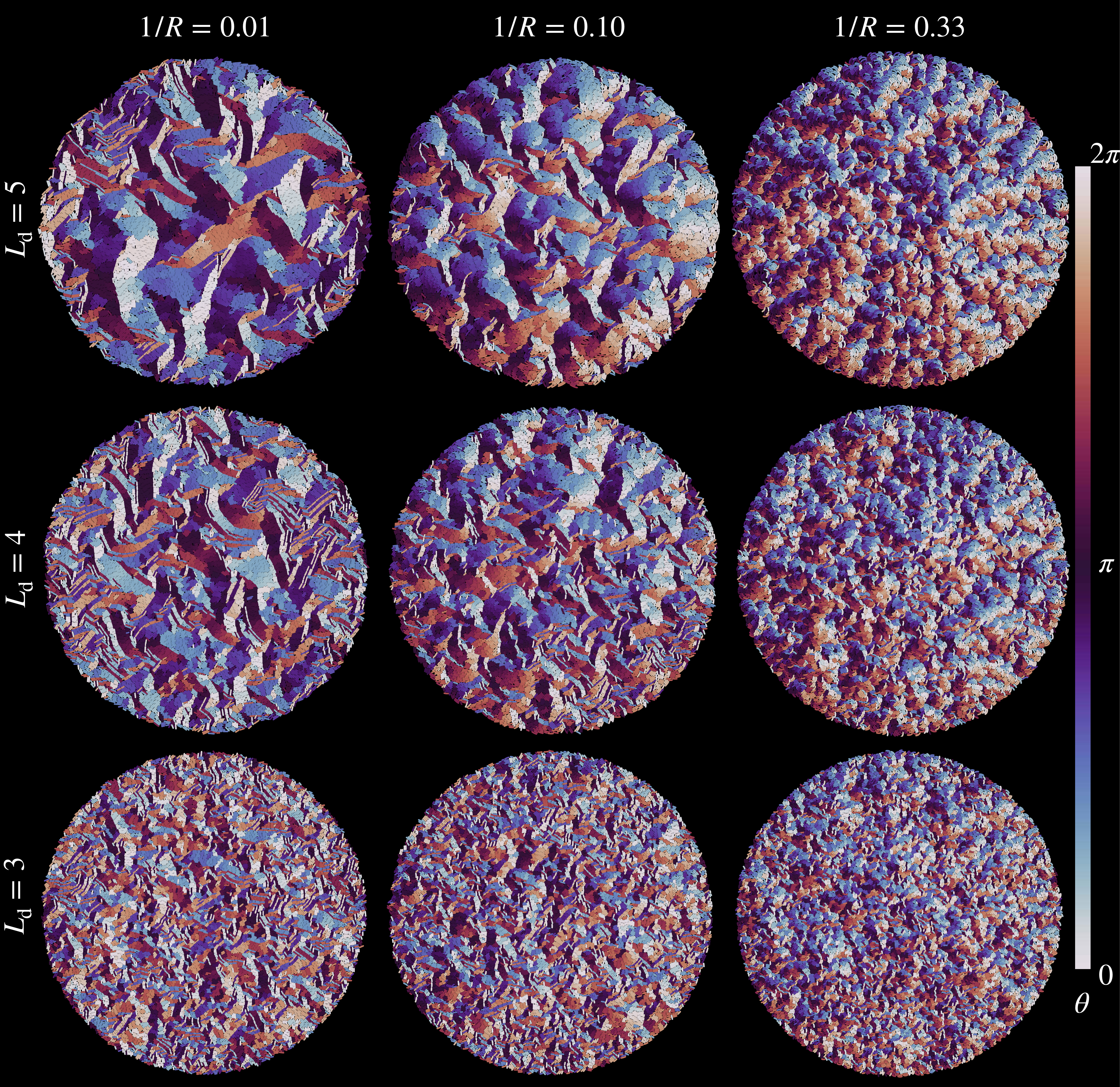}
  \caption{\label{fig2} Effects of curvature $1/R$ of the growing individual cells and their threshold division length $L_\RM{d}$ on the resulting overall structure of the colony. From left to right, curvature increases as $1/R=0.01$, $0.10$, and $0.33$. From top to bottom, the threshold division length decreases as $L_\RM{d}=5$, $4$, and $3$.
  Increasing curvature and decreasing threshold division length lead to finer, less ordered structures and decreasing average domain size. 
  Color represents the angle of orientation $\theta_i\in[0,2\pi)$ of each cell $i$, see Fig.~\ref{fig1}(a), as quantified by the color bar on the right-hand side.
  The mean growth rate is set to $g=5\times10^{-5}$.
  }
\end{figure*}
\begin{figure*}[t]
\centering
  \includegraphics[width=17cm]{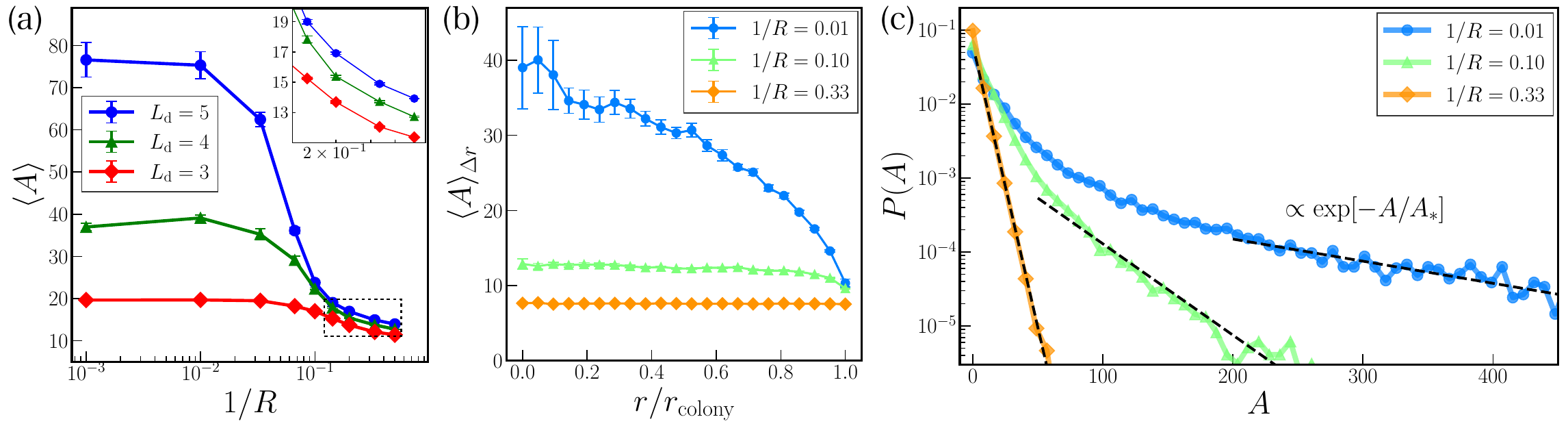}
  \caption{\label{fig3}(a) Average domain area $\expval{A}$ as a function of cellular curvature $1/R$ for threshold division lengths $L_\RM{d}=5$, $4$, and $3$. 
  The average is taken over all domains located within the inner circle of half the colony radius $r_\RM{colony}$. 
  In the inset we provide a magnification of the regime for large curvature $1/R$ (the dotted rectangle).
  (b) Dependence of the average domain area for different cellular curvatures $1/R=0.01$,$0.10$, and 
 $0.33$ on the distance $r$ from the center of the colony. The threshold division length is set to $L_\RM{d} = 5$. We calculate the average $\expval{A}_{\Delta r}$ for each value of $r$ from all domains located within an annular ring 
 of width $\Delta r = 5$.
 Typically, $r_\RM{colony}\sim 100$ at times of evaluation. 
 (c)~Probability distribution $P(A)$ of the domain area $A$ within colonies of cellular curvature $1/R=0.01$, $0.10$, and $0.33$. Dashed lines represent exponential behavior $P(A)\sim \exp[-A/A_*]$ within the indicated regimes. From corresponding fits, we obtain values of characteristic domain areas $A_*=145.4$, $35.2$, and $5.7$, respectively, which decrease with increasing curvature. We set the threshold division length to  $L_\RM{d} = 5$. 
  The average is taken over 30 samples.
  }
\end{figure*}
As outlined above, straight-growing microorganisms are frequently represented by growing spherocylinders that divide into two halves after reaching a certain threshold length~\cite{you2018geometry}. 
To now consider non-straight microorganisms, we add curvature to these spherocylinders. 
For this purpose, each curved cell is described as an arc (part of a circle) of radius $R$ and thickness $d_0$, see \fig{fig1}(a).
The length of the $i$-th cell is given by the arc length $L_i=R\phi_i$, where $\phi_i$ is the opening angle of the segment of the disk containing the microorganism at its bounding arc. 
Figures~\ref{fig1}(c), (d), and (e) show illustrations of curved cells of curvature $1/R=0.01, 0.10,$ and $0.33$.

We consider the growth of a flat colony or film in the two-dimensional $x$-$y$ plane. 
Growth of the $i$-th cell is represented by a continuously increasing arc length $L_i$. Here, we assume $L_i$ to increase linearly in time $t$, 
\begin{equation}
\dv{L_i(t)}{t} = G_i, \label{model_1}
\end{equation}
where $G_i$ is the constant growth rate chosen independently for each cell from the uniform distribution $G_i\in[g/2,3g/2]$. 
The parameter $g$ stands for the average growth rate. 
We assume that each cell grows while keeping the curvature constant.  

Each cell divides into two cells of equal size when its arc length reaches the threshold value $L_\RM{d}$. The growth rates for the two resulting cells after each division are again selected randomly from $[g/2,3g/2]$. 
Initially, we place a single curved cell, from where the colony develops, while the total number of microorganisms $N(t)$ increases in time. 

During the entire process of growth, we need to evaluate the mutual steric interactions resulting from volume exclusion between the cells and their consequences. To this end, at each time step, we find, for any two cells $i$ and $j$ in contact, the two inner tangent circles to the curved spherocylinders that are nearest to each other~\cite{ghosh2015mechanically}, see \fig{fig1}(b). The center-to-center vector between these two circles is termed $\bm d_{ij}$. 
We assume the force of mutual interaction $\bm F_{ij}$ between the two circles to be described by the Hertzian elastic model~\cite{landau1986theory}.
The force $\bm F_{ij}$ is then mapped to the interaction between the $i$-th and $j$-th cells. 

We consider the dynamics of each cell $i$ to be overdamped and to follow the equations of motion~\cite{you2018geometry}
\begin{align}
\dv{\bm r_i(t)}{t} &= \frac{1}{\zeta L_i(t)}\sum_{j=1}^{N(t)}\bm F_{ij}(t),\label{model_3} \\
\dv{\theta_i(t)}{t} &= \frac{12}{\zeta L^3_i(t)}\sum_{j=1}^{N(t)}\qty[\bm \ell_{ij}(t) \times \bm F_{ij}(t)]\cdot \hat{\bm e}_z. \label{model_4}
\end{align}
Here, $\bm r_i(t)$ denotes the position of the center of mass of the $i$-th cell, and $\theta_i(t)$ quantifies the orientation of the $i$-th cell, both at time $t$.
The direction of the orientation of the $i$-th cell $\hat{\bm e}_i =(\cos\theta_i,\sin\theta_i)$ is set from the center of the arc to the center-of-mass position of the cell, normal to the arc length, see \fig{fig1}(a).
$\theta_i$ takes values in the range $[0,2\pi)$ for curved cells, as curvature provides a sense.  
$\zeta$ sets the strength of the friction with the substrate per arc length of the cell and is supposed to be identical for all microorganisms. 
We remark that we assume the friction to be isotropic to concentrate on the steric effects of curvature, while the effect of anisotropic friction was taken into account in previous studies on straight cells~\cite{acemel2018computer, Doumic2020, Shimaya2022PNAS_Nexus}. 
Moreover, $\bm \ell_{ij}(t)$ is the vector from the center-of-mass position to the point of the cell where the force acts. 
$\hat{\bm e}_z $ is the unit vector pointing into the $z$-direction, normal to the plane. 

To simulate Eqs.~(\ref{model_1}), (\ref{model_3}), and (\ref{model_4}) we use the semi-implicit Euler method with a time step of $\Delta t=1\times10^{-2}$.
In Eq.~(\ref{model_3}), the force $\bm F_{ij}$ is given by the Hertzian force~\cite{landau1986theory}:
\begin{equation}
\bm F_{ij} = Ed_0^2\qty(1-\frac{d_{ij}}{d_0})^{3/2} \frac{\bm d_{ij}}{d_{ij}} \Theta(d_0 - d_{ij}) \label{model_2}.
\end{equation}
Here, $E$ is an effective elastic modulus and $\Theta(\cdot)$ represents the Heaviside step function. 
We choose $\zeta/E$ and $d_0$ as the units of time and length, respectively. 
The control parameters in the simulation are the dimensionless average growth rate $\zeta g/d_0 E$, the maximum length of a cell $L_\RM{d}/d_0$, and the radius (inverse curvature) $R/d_0$.
We denote $g$, $L_\RM{d}$, and $R$ in these units as dimensionless parameters unless noted otherwise.
The average growth rate is fixed to be $g=5\times 10^{-5}$, and we vary $L_\RM{d}$ and $R$. 
We terminate the simulations when $L_\RM{tot} = \sum_{i=1}^{N(t)}(L_i(t) + d_0)\geq 3.75\times 10^4$. 

\section*{Results}
\subsection*{Characteristics of the domain structure resulting from colony growth}
\begin{figure*}[t]
\centering
  \includegraphics[width=17cm]{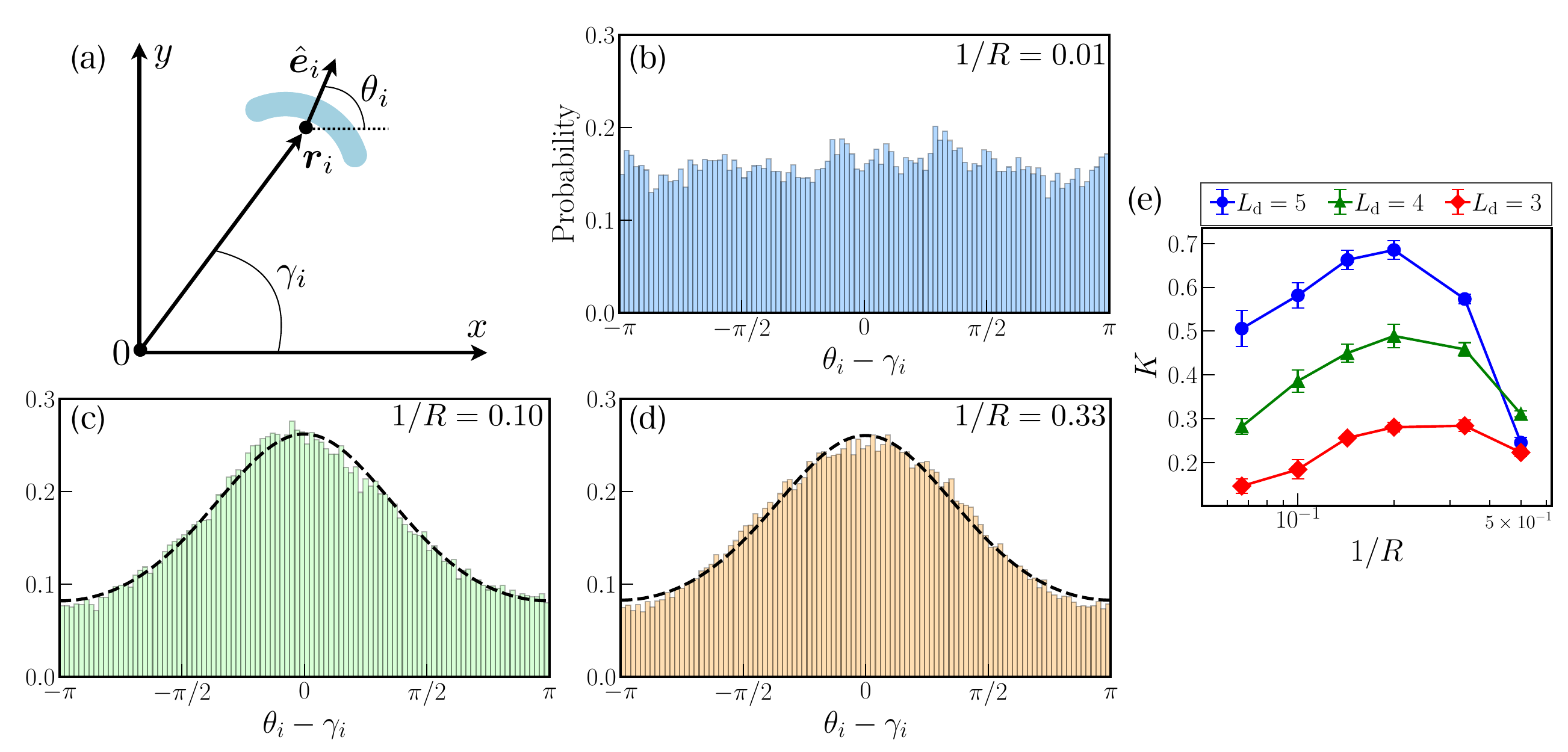}
  \caption{\label{fig5} (a) Schematic illustration of evaluating the spatio-orientational coupling. We determine for each cell $i$ the angle $\gamma_i$ for the radial direction running from the center of mass of the colony through the center of mass of the cell $\bm{r}_i$. 
  This angle is compared to the individual angular orientation $\theta_i$ for each cell. 
  Histograms of $\theta_i-\gamma_i$ for cellular curvatures (b)~$1/R=0.01$, (c)~$1/R=0.10$, and (d)~$1/R=0.33$ are depicted for the threshold division length $L_\RM{d}=5$. 
  In (c) and (d), the dashed lines are obtained by fitting the circular normal distribution $\exp[K\cos(\theta-\gamma)]/(2\pi I_0(K))$ to the histograms. 
  The parameter $K$ resulting from the fits quantifies the peakedness of the distribution and is depicted in panel (e) as a function of cellular curvature $1/R$ for the different threshold division lengths $L_\RM{d}=5$, $4$, and $3$.
  }
\end{figure*}
On this basis, we now explore the fundamental effects of growing and dividing curved microorganisms on the resulting film structures. 
Videos of the simulated growth process starting from a single, randomly oriented cell under the aforementioned conditions 
are available in the supporting material.
Figure~\ref{fig1}(f) shows a typical resulting time evolution of the total number of cells $N(t)$ in a growing colony. It confirms the expected exponential increase. For illustration, Fig.~\ref{fig1}(g) provides three intermediate snapshots.
Final example configurations upon variation of the basic parameters, that is, curvature $1/R$ [inverse radius, see Fig.~\ref{fig1}(a)] and threshold division length $L_\RM{d}$, are depicted in Fig.~\ref{fig2}.
All quantities described in the following are analyzed in the final stages of the simulations. 

In Fig.~\ref{fig2}, color represents the angle of the orientational vector $\theta_i\in[0,2\pi)$  of the $i$-th cell, see Fig.~\ref{fig1}(a). 
The columns from left to right show increasing cellular curvature $1/R$. In the left-most column, for the smallest curvature $1/R=0.01$, the individual cells are basically straight, see \fig{fig1}(c), and we qualitatively expect corresponding behavior. 
As already reported in Ref.~\cite{you2018geometry} for straight cells, one does not find global orientational order. Instead, locally aligned domains form a kind of ``mosaic'' structure. 
These mesoscopic domains decrease in size as the division length $L_\RM{d}$ decreases. We recover this trend from top to bottom when we reduce this length.  

Our central observation and message is that increasing curvature $1/R$ and decreasing threshold division length $L_\RM{d}$ generally and substantially reduce the average size of the mesoscopic domains and thus the extent of local alignment. This effect becomes apparent from Fig.~\ref{fig2} when turning from left to right (curvature $1/R$ increasing as $0.01$, $0.10$, $0.33$) and from top to bottom (threshold division length $L_\RM{d}$ decreasing as $5$, $4$, $3$). 
Both trends express that straight elongated objects support orientational alignment. Variations by reducing the aspect ratio or deviating from the straight shape lead to less pronounced orientational alignment.

In the following, we analyze these effects more quantitatively. 
To this end, we determine the average area $\langle A\rangle$ of the mesoscopic domains at any given state of the colony.
To determine the sizes of the domains of orientational order, we consider two neighboring cells $i$ and $j$ as belonging to the same mesoscopic domain, if they contact each other and show approximately identical orientation. The first criterion is satisfied if the force in Eq.~\ref{model_2} becomes nonzero.
The second criterion applies if their orientation angles fulfill $\abs{\theta_i - \theta_j}<0.1$.
(The choice of the precise value of this upper bound for $\abs{\theta_i - \theta_j}$ does not affect our results qualitatively, see the supporting material).
The area $A$ of any resulting mesoscopic domain is calculated as the areal sum over all cells within that domain. 
We evaluate $\langle A\rangle$ as a function of cellular curvature.
Ensemble averages are taken over $10$ different simulations.
Figure~\ref{fig3}(a) shows the average domain area $\expval{A}$ as a function of the cellular curvature $1/R$ for the different threshold division lengths $L_\RM{d}=5,4,$ and $3$. 
To exclude surface effects, 
$\expval{A}$ is calculated in each case for the interior circular film area of half the colony radius $r_\RM{colony}/2$. Here, the radius of the entire colony is defined via $r_\RM{colony}(t) = \max_{i=1,...,N}|\bm r_i(t)|$. The origin is set by the position of the initial cell. 
In our case, $r_\RM{colony}\approx 100$, but the precise value slightly depends on the curvature and division length (see the supporting material).
From Fig.~\ref{fig3}(a), we find that the average domain area $\expval{A}$ is almost constant when increasing the curvature of the cells $1/R$ at small values, that is, for basically straight microorganisms. Yet, from a certain value of $1/R$ onwards, $\expval{A}$ significantly decreases. 
This effect is most pronounced for $L_\RM{d} = 5$, while for $L_\RM{d} = 4$ and $3$ it is more moderate. 
Since initially the averaged domain size $\langle A\rangle$ is larger for straight cells of $L_\RM{d}=5$ than for $L_\RM{d}=4$ and $3$, also the reduction in this value can be larger, as is the case in Fig.~\ref{fig3}(a). Still, the magnitudes of  $\langle A\rangle$ remain in the order $L_\RM{d}=5$, $4$, and $3$ from top to bottom. Our observations are in line with the snapshots displayed in Fig.~\ref{fig2}. There, we observe by eye reduction in domain size from top to bottom and from left to right in each column and line. 

Next, we address in Fig.~\ref{fig3}(b) how the area of the domains depends on their location within the colony, together with the effect of cellular curvature on this dependence. 
We slice the entire colony into concentric, annular rings of width $\Delta r$ for quantification. Their location is given by their distance $r$ from the center of the colony.
Figure~\ref{fig3}(b) shows the averaged domain area within each ring, $\expval{A}_{\Delta r}$, as a function of $r$.
The ensemble average is taken over 30 different simulations.
For basically straight cells of curvature $1/R=0.01$, we observe a decreasing trend of $\expval{A}_{\Delta r}$ towards the outer rim of the colony, in line with previous reports in Ref.~\cite{you2018geometry}. 
Instead, for the increasingly curved cases of $1/R=0.10$ and $0.33$, $\expval{A}_{\Delta r}$ remains rather constant towards the outer rim. In line with our results above, we infer that increased curvature alone already implies a significant reduction in domain area and, thus, an increase in mesoscopic disorder. 
For straight cells, reducing the division length 
$L_\RM{d}$ has similar effects \cite{you2018geometry}. It likewise smoothens the distribution of domain area from the center towards the outer rim of the colony. 

Additionally, we evaluate the probability distribution $P(A)$ for the observed mesoscopic domain areas $A$ in a given colony. 
The distribution shows exponential behavior for straight cells $P(A)\sim \exp[-A/A_*]$, where $A_*$ represents a characteristic domain area. $A_*$ decreases together with the division length $L_\RM{d}$~\cite{you2018geometry, Lama2024PNAS}.
We infer from Fig.~\ref{fig3}(c) that the exponential dependence can still be observed with increasing cellular curvature $1/R$. 
However, increasing curvature reduces the characteristic area $A_*$, in line with our qualitative observation in Fig.~\ref{fig2}.

\subsection*{Spatio-orientational coupling in colonies of curved cells}
\begin{figure*}[t]
  \centering
  \includegraphics[width=17.5cm]{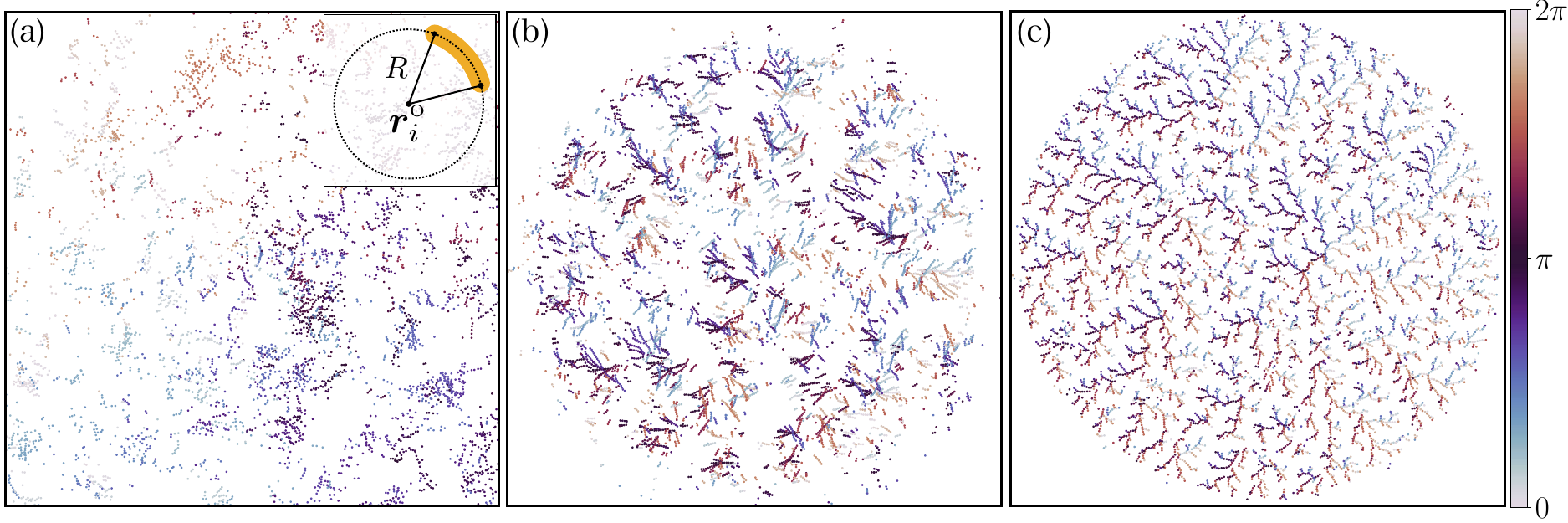}
  \caption{\label{fig6} Center-of-arc positions $\bm r_i^\RM{o}$ of the microorganisms in the film, see the inset of panel (a), together with color coding of the individual orientations for increasing cellular curvature (a)~$1/R=0.01$, (b)~$1/R=0.10$, and (c)~$1/R=0.33$. The threshold division length is set to $L_\RM{d}=5$ so that the illustrations reconsider the state of the colonies depicted in the top row of \fig{fig2}. For the almost straight cells in (a), the radii of the arcs are correspondingly large, therefore several center-of-arc positions $\bm r_i^\RM{o}$ are located outside the box. As curvature $1/R$ increases from left to right, the mutual arrangements of $\bm r_i^\RM{o}$ transition from (a) rather scattered, gas-like to (c) organized, branched filamentous structures. 
  The latter indicates the stacked organization of curved cells in these colonies. 
  }
\end{figure*}
Thus, as demonstrated above, the curvature of individual cells can have an effect qualitatively similar to reduced division lengths. At least, this concerns the average mesoscopic domain size in the colonies. 
A similar tendency of reduced domain size could be provoked by reduced aspect ratio of the individual cells by increasing their thickness. 
Yet, additionally, new structural consequences of cellular curvature emerge that are not present in colonies of straight cells. A specific example is the coupling of cellular orientation to the position within the colony. Through these measures, we can qualitatively and quantitatively distinguish the appearance of colonies of straight cells of low division length from films of curved cells of longer division length, which by eye appear relatively similar; see the bottom left and top right in Fig.~\ref{fig2}. 

Overall, with increasing cellular curvature, the microorganisms tend to align themselves tangentially to concentric rings around the center of the colony. 
Another way of formulating this trend is that they rotate the center positions of their contour arcs towards the center of the colony. 
Yet another interpretation is that the individual orientation of their contours mimics the overall circular shape of the colony. 

To illustrate this trend, we evaluate their spatio-orientational coupling. 
We determine for each cell the angle $\gamma_i$ of the line running from the center of the colony to the center of mass of the cell. 
That is, $\gamma_i$ measures the radial direction of the colony at the position of the considered cell. 
We compare this angle with the angular orientation $\theta_i$ of each cell; see Fig.~\ref{fig5}(a). 
Proceeding along these lines for all cells, we calculate histograms as a function of $\theta_i-\gamma_i$. 
The trend with increasing curvature $1/R=0.01$, $0.10$, to $0.33$ is illustrated in Fig.~\ref{fig5}(b)--(d). 
The threshold division length here is set to $L_\RM{d}=5$. 
Ensemble averages over ten colonies are performed for each curvature. 

We observe in Fig.~\ref{fig5}(b) that the distribution is flat for the nearly straight cells of curvature $1/R=0.01$. Thus, we do not find spatio-orientational coupling for non-curved cells. 
However, increasing cellular curvature to $1/R=0.10$ and $0.33$, a broad peak emerges in the distribution around $\theta_i=\gamma_i$, see Figs.~\ref{fig5}(c) and (d).
Consequently, here, the spatio-orientational coupling emerges as described above. 
This trend is observed even for the smaller division length of $L_\RM{d} = 3$ and $4$ (see the supporting material). 

To further quantify the degree of spatio-orientational coupling, we fit the circular normal distribution (also known as the von Mises distribution), which is a $2\pi$-periodic analog of the normal distribution~\cite{mardia_directional}, to each of the unimodal histograms:
\begin{equation}
    f(\theta) = \frac{1}{2\pi I_0(K)}e^{K\cos(\theta-\gamma)},
\end{equation}
where $I_0(K)$ is the zeroth-order modified Bessel function of the first kind. The dashed lines in Figs.~\ref{fig5}(c) and (d) confirm the suitability of the fitting function. 

The parameter $K$ characterizes the strength of the modulation and thus the height of the peak of the distribution. 
Therefore, the larger the value of $K$, the more pronounced the spatio-orientational coupling is. 
Figure~\ref{fig5}(e) displays the resulting dependence of $K$ on the curvature $1/R$ for $1/R\geq 0.067$. 
The error bars indicating standard deviations in Fig.~\ref{fig5}(e) were obtained from the fits to the different realizations of the system, averages of which are displayed as the data points. 
We again consider threshold division lengths of $L_\RM{d} = 5$, $4$, and $3$.

All three lines feature a maximum around $1/R=0.2$. 
Elevated values of $K$ imply more strongly peaked distributions. 
This suggests that the most pronounced spatio-orientational coupling occurs at intermediate curvature. 
It implies that the normals to the arc-like contours of the microorganisms tend to point radially outward of the colony. 
When reducing curvature $1/R$ from the maximum of the curves, $K$ drops, signaling flatter distributions and thus lower spatio-orientational coupling. This trend is in line with observations for straight cells as in Fig.~\ref{fig5}(b), where spatio-orientational coupling is absent. Thus, the effect of spatio-orientational coupling is introduced by curvature. 
As for the decrease of $K$ at large curvatures, possibly, this is due to the higher isotropy of strongly bent cells. 
In other words, the contour of a strongly curved long cell closes into a circle, which implies more isotropic steric interactions. 

\subsection*{Specific substructures in colonies of curved cells}
Finally, the structural elements in colonies of curved microorganisms are qualitatively different from those of straight cells. 
We noted above the spatio-orientational coupling on an individual level. Here, we focus on mesoscopic structural elements consisting of several cells. 

First, from snapshots of the colonies consisting of curved cells, see \fig{fig2}, we infer by close inspection the formation of scattered ring-like structural elements. This type of arrangement, by construction, is not possible for straight cells. However, it does not seem to be dominant in the colony. 
Much more obviously, stacked arrangements of curved cells emerge. This is in contrast to the rather patch-like arrangements of the mesoscopic domains of straight cells. The spatial organization in stacks is triggered and supported when individual arc-like cells are piled into each other. 

To illustrate this feature and to contrast the organization in colonies of straight and curved cells, we include Fig.~\ref{fig6}. 
There, we plot by dots the center positions $\bm r_i^\RM{o}$ of the arcs of all cells, see \fig{fig1}(a) and the inset of Fig.~\ref{fig6}(a), together with the color coding representing cellular orientations. 
For almost straight microorganisms of curvature $1/R=0.01$, we hardly observe any such stacking of nearby aligned cells; see \fig{fig6}(a). The spatial organization of the center-of-arc positions $\bm r_i^\RM{o}$ is relatively homogeneous and gas-like. 
However, the situation changes qualitatively with increasing curvature, here to $1/R=0.10$ and $0.33$ in Figs.~\ref{fig6}(b) and (c). 
With increasing curvature, we find branched filamentous elements of $\bm r_i^\RM{o}$. 
They are a clear signature of the line-like stacking arrangements of neighboring curved cells in the corresponding colonies, in contrast to what is observed for straight cells. 
Curved cells can be stacked like cups into each other to save space, whereas this is not necessary for straight objects that can form orientationally ordered domains for space-efficient arrangement.

\section*{Conclusion}
In summary, we address the planar growth of colonies of bent bacteria and other microorganisms. To this end, we consider persistent curvature of the individual units. These units grow along their arc length. When reaching a certain threshold length, they divide. Only mutual steric interactions that represent volume exclusion between the individuals are included. 

We analyze how the curvature of the individual units affects the structure of the growing colonies. Notably, the typical size of orientationally ordered domains substantially reduces with increasing cellular curvature. 
While the effects of curvature on domain size are qualitatively similar to those of decreasing threshold division length, different structures emerge when compared to the case of straight microorganisms.
One difference becomes apparent by the correlation between the orientation of the cells and their position in the film, as curved cells tend to orient concentrically to the colony. Another feature is the emergence of branched filamentous structures of strings of stacked curved microorganisms. 

Our focus in this initial study is on the fundamental phenomenology underlying and resulting from the expansion of colonies composed of growing and dividing curved microorganisms. 
We concentrate on the generic effect of volume exclusion.
On this basis, extensions into various directions are conceivable. 
For instance,  mechanical regulation of the speed of individual growth~\cite{winkle2017modeling, wittmann2023collective}, diffusive processes~\cite{acemel2018computer}, polydispersity of aspect ratios~\cite{Berg2024}, or anisotropic friction~\cite{acemel2018computer, Doumic2020, Shimaya2022PNAS_Nexus} can become important.
Additional types of mutual interaction can come into play, for example, adhesional forces~\cite{jin2020influence}. Moreover, beyond planar arrangements in monolayered colonies, the transition to bi- and multilayered structures and thus into the third dimension becomes important in reality at some point~\cite{beroz2018verticalization, you2019mono, dhar2022self}. 

Besides, one may think not only about extensions that concern the influence of properties related to the cells themselves. 
For instance, the role of nutrient supply and its effect on the growth rate represents another interesting aspect and takes the impact of the environment into account~\cite{ghosh2015mechanically, farrell2017mechanical}. 
Along these lines, the mechanical properties of the substrate can play a notable role as well. Soft substrates deform under the action of the stress induced by living biological cells~\cite{sabass2008high, tanimoto2014simple}. Resulting interactions via deformations of elastic films are long-ranged. They can affect or induce mutual orientational order~\cite{schwarz2002elastic, bischofs2004elastic}. 

Many of the just-listed aspects are being addressed for straight cells. Clarifying the role of curvature of the individuals in combination with all these effects implies an interesting subject of research on its own. For instance, the shape of the individual curved objects becomes increasingly isotropic during growth towards a straight circle. In this way, during the growth cycle of each individual, anisotropy in friction and in mutual long-ranged interaction with and through the substrate may vary. 

\section*{Author Contributions}
A.M.M.\ designed the research; Y.K.\ performed the simulations and analyzed the numerical data; T.K.\ and A.M.M.\ assisted with the numerical implementation and data analysis; all authors discussed and interpreted the results; all authors wrote the paper.

\section*{Acknowledgments}
We thank the Deutsche Forschungsgemeinschaft (German Research Foundation, DFG) for support through the Heisenberg grant no.\ ME 3571/4-1. A.M.M.\ further thanks the DFG for support through the research grant ME 3571/12-1. 
Y.K.\ acknowledges support from JSPS KAKENHI (grant no.\ JP23KJ1068). T.K.\ acknowledges support by the JST FOREST Program (grant no.\ JPMJFR212T), AMED Moonshot Program (grant no.\ JP22zf0127009), and JSPS KAKENHI (grant no.\ JP24H02203).


\section*{Supporting Material}
Supporting material is attached below.
%
\ifarXiv
    

\section*{Supplementary material}

# Supporting Information for "Effects of curvature on growing films of microorganisms"

Yuta Kuroda

*Institut für Physik, Otto-von-Guericke-Universität Magdeburg,*
*Universitätsplatz 2, 39106 Magdeburg, Germany and*
*Department of Physics, Nagoya University, Nagoya 464-8602, Japan*

Takeshi Kawasaki

*Department of Physics, Nagoya University, Nagoya 464-8602, Japan*

Andreas M. Menzel

*Institut für Physik, Otto-von-Guericke-Universität Magdeburg, Universitätsplatz 2, 39106 Magdeburg, Germany*
(Dated: August 21, 2024)

## 1. DEPENDENCE OF THE RADIUS OF THE COLONY ON THE CELLULAR CURVATURE AND THE THRESHOLD DIVISION LENGTH

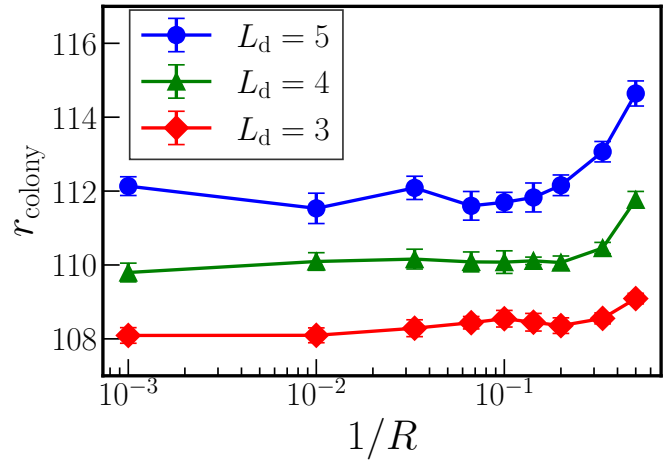

FIG. S1. Radius of the colony $r_{\text{colony}}$ as a function of the curvature of the individual cells $1/R$ for different threshold division lengths $L_{\text{d}}$.

Figure S1 depicts the radius of the entire colony $r_{\text{colony}}$ for our final configurations at which we stop our simulations as a function of the curvature of the individual cells $1/R$. Results are displayed for the three division lengths $L_{\text{d}} = 5$, 4, and 3. Ensemble averages are taken over 10 different samples for each data point. For all values of $L_{\text{d}}$, the radius $r_{\text{colony}}$ increases as the curvature increases for elevated values of $1/R$. This is natural as the colony of curved cells contains more gaps compared to the situation of straight constituents.

## 2. SPATIO-ORIENTATIONAL COUPLING FOR THRESHOLD DIVISION LENGTHS $L_{\text{d}} = 3$ AND 4

In the main text, we show the histograms for $\theta_i - \gamma_i$ as a measure of spatio-orientational coupling for the threshold division length $L_{\text{d}} = 5$. This is the largest threshold division length considered in our simulations, see Fig. 4 in the main text. $\theta_i$ measures the individual angular orientation of each cell $i$, while $\gamma_i$ quantifies its angular position within the colony, see Fig. 4(a) in the main text. Here, we include in Fig. S2 the corresponding histograms of $\theta_i - \gamma_i$ for the smaller division lengths $L_{\text{d}} = 3$ and 4. Obviously, even for these, we find inhomogeneous unimodal distributions for the cellular curvatures $1/R = 0.10$ and 0.33. Thus, again spatio-orientational coupling is observed. However, the widths of the distributions as quantified by the standard deviation $\sigma$ are larger for the smaller threshold division lengths $L_{\text{d}} = 3$ and 4.

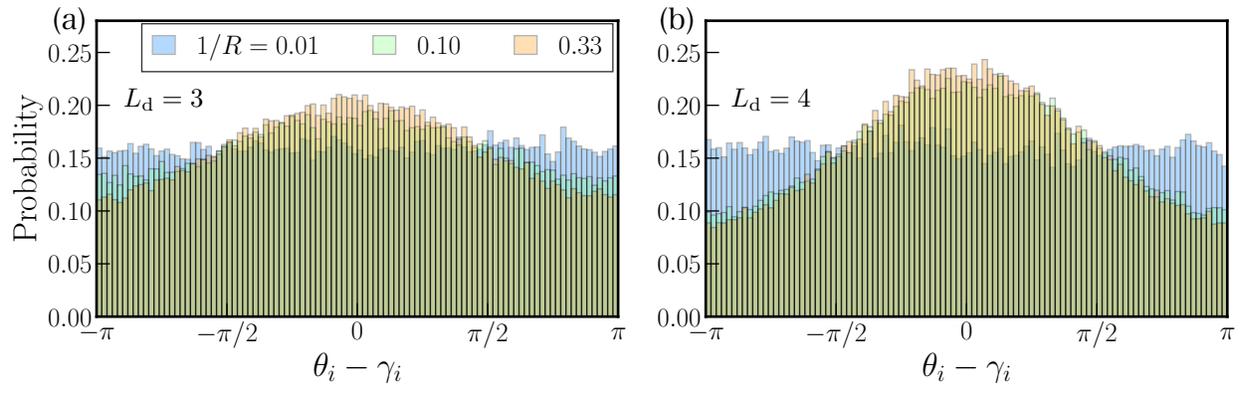

FIG. S2. Histograms of $\theta_i - \gamma_i$ as a measure for spatio-orientational coupling within the colony for threshold division lengths (a) $L_d = 3$ and (b) $L_d = 4$.


\fi
\end{document}